\documentclass{iau_FM}
\usepackage{graphicx}
\usepackage{comment}
\usepackage{url}

\title[Hunting for stellar coronal mass ejections] 
{Hunting for stellar coronal mass ejections}

\author[Kosuke Namekata et al.]   
{Kosuke Namekata$^1$,
Hiroyuki Maehara$^2$,
Satoshi Honda$^3$,
Yuta Notsu$^{4,5,6}$,
Daisaku Nogami$^7$,
\and Kazunari Shibata$^8$}

\affiliation{$^1$ALMA Project, NAOJ, NINS, Osawa, Mitaka, Tokyo, 181-8588, Japan \\ email: {\tt namekata@kusastro.kyoto-u.ac.jp, kosuke.namekata@astro.nao.ac.jp} \\[\affilskip]
$^2$Okayama Branch Office, Subaru Telescope, NAOJ, NINS, Kamogata, Asakuchi, Okayama 719-0232, Japan\\[\affilskip]
$^3$Nishi-Harima Astronomical Observatory, Center for Astronomy,University of Hyogo, Sayo, Hyogo 679-5313, Japan\\[\affilskip]
$^4$Laboratory for Atmospheric and Space Physics, University of Colorado Boulder, 3665 Discovery Drive, Boulder, CO 80303, USA\\[\affilskip]
$^5$National Solar Observatory, 3665 Discovery Drive, Boulder, CO 80303, USA\\[\affilskip]
$^6$Department of Earth and Planetary Sciences, Tokyo Institute of Technology, 2-12-1 Ookayama, Meguro-ku, Tokyo 152-8551, Japan\\[\affilskip]
$^7$Astronomical Observatory, Kyoto University, Sakyo, Kyoto 606-8502, Japan\\[\affilskip]
$^8$Kwasan Observatory, Kyoto University, Yamashina, Kyoto 607-8471, Japan\\[\affilskip]
}

\pubyear{2022}
\jname{Astronomy in Focus} 
\editors{H. Korhonen, J. Espinosa \& M. Smith-Spanier eds.}
\begin{document}

\maketitle

\begin{abstract}
Solar flares are often accompanied by filament/prominence eruptions, sometimes leading to coronal mass ejections (CMEs). By analogy, we expect that stellar flares are also associated with stellar CMEs whose properties are essential to know the impact on exoplanet habitability. Probable detections of stellar CMEs are still rare, but in this decade, there are several reports that (super-)flares on M/K-dwarfs and evolved stars sometimes show blue-shifted optical/UV/X-ray emissions lines, XUV/FUV dimming, and radio bursts. Some of them are interpreted as indirect evidence of stellar prominence eruptions/CMEs on cool stars. More recently, evidence of stellar filament eruption, probably leading to a CME, is reported even on a young solar-type star (G-dwarf) as a blue-shifted absorption of H$\alpha$ line associated with a superflare. 
Notably, the erupted masses for superflares are larger than those of the largest solar CMEs, indicating severe influence on exoplanet environments. The ratio of the kinetic energy of stellar CMEs to flare energy is significantly smaller than expected from the solar scaling relation and this discrepancy is still in debate. We will review the recent updates of stellar CME studies and discuss the future direction in this paper.

\keywords{stars: late-type, stars: magnetic fields, stars:activity, Sun: flares, stars:flare, Sun: coronal mass ejections (CMEs), stars: mass loss}
\end{abstract}

\firstsection 
\section{Introduction and motivation}

Solar and stellar flares are explosive phenomena on the surface and they have been observed in the wavelength from X-rays to radio.
They are thought to be caused by the conversion of magnetic energy into kinetic and thermal energy via magnetic reconnection \cite[(e.g., Shibata \& Magara 2011; Namekata et al. 2017)]{2011LRSP....8....6S; 2017ApJ...851...91N}.
Solar flares are sometimes associated with filament/prominence eruptions and  coronal mass ejections (CMEs; \cite[Gopalswamy et al. 2003]{2003ApJ...586..562G}).
The CMEs often generate shockwaves and high-energy particles which are observed by the artificial satellites.
Moreover, active regions and solar flares produce strong X-ray and extreme ultraviolet (EUV) radiations which have impacts on the planetary ionosphere.
This is how the magnetic activities on the Sun have affected planetary habitability and human civilization in the solar system \cite[(Cliver et al. 2022)]{2022LRSP...19....2C}.

Recent Kepler and TESS observations revealed that F, G, K, and M dwarfs frequently host exoplanets. 
The exoplanets around the active stars are expected to be subject to eruptive flares, high XUV radiation, and stellar winds that may cause significant changes in their atmospheric loss and chemistry \cite[(e.g., Linsky 2019; Airapetian et al. 2020)]{2019LNP...955.....L,2020IJAsB..19..136A}. 
It is possible that the magnetized plasma of stellar CME may strip the atmosphere of the close-in exoplanet. 
It is also proposed that high-energy particles produced by stellar CMEs can cause chemical reactions in planetary atmospheres and produce molecules important for habitability and the origin of life, such as greenhouse gases and prebiotic chemistry \cite[(Airapetian et al. 2016)]{2016NatGe...9..452A}.
Therefore, finding and characterizing stellar CMEs have been more and more highlighted for understanding exoplanetary habitability.

Although stellar flares have long been observed, the Kepler and TESS mission has dramatically improved our statistical understandings of the relation between stellar parameters (e.g., surface temperature, rotation period, and age) and the nature of flares (e.g., frequency of occurrence, energy, etc.) over the last decade \cite[(e.g., Davenport 2016; Maehara et al. 2012)]{2016ApJ...829...23D,2012Natur.485..478M} .
As the stellar age increases, the rotation velocity slows down due to magnetic breaking, and as a result the flare frequency and maximum energy are found to be several orders of magnitude smaller \cite[(e.g., Maehara et al. 2012; Notsu et al. 2019; Okamoto et al. 2021)]{2012Natur.485..478M,2019ApJ...876...58N,2021ApJ...906...72O}.
This means that young stars may also have a greater impact on planetary environments, where habitable environments are being formed.
Also, Kepler and TESS found that cooler stars (e.g., M-dwarfs) having close-in habitable zone generally show much higher flare activity than solar-type stars \cite[(e.g., Davenport 2016; Maehara et al. 2021)]{2016ApJ...829...23D,2021PASJ...73...44M}. 
These discoveries have suggested that the possible habitable worlds on close-in exoplanets around cool stars can experience more severe effect of magnetic activities of central stars than human beings have experienced on the Earth.

\section{Do stellar CMEs occur? -- Hints from solar observations --}

\begin{figure}[htb]
\begin{center}
 \includegraphics[width=5.in]{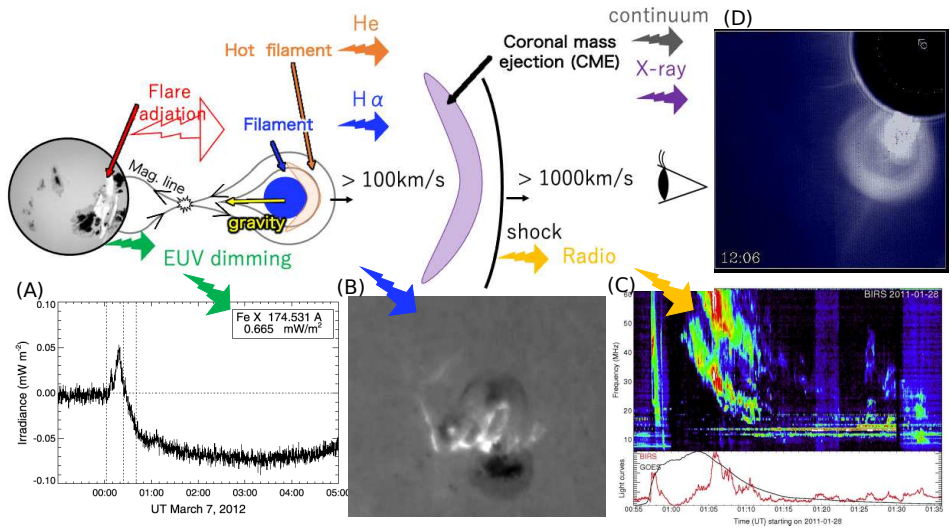} 
 \caption{Schematic picture of possible ways to detect stellar mass ejection on the basis of solar observations (reproduced from Namekata 2021). (A) Post-flare EUV dimming on the stellar surface as an indirect signature of mass ejection \cite[(Harra et al. 2016; Courtesy of Harra et al.)]{2016SoPh..291.1761H}. (B) Prominence eruption (Courtesy of Hida Observatory). (C) Type-II radio burst as an evidence of CME shock propagating through the interplanetary space \cite[(Crosley et al. 2017; Courtesy of Crosley et al.)]{2017ApJ...845...67C}. (D) CME observed by white-light coronagraph (Courtesy of HAO/SMM C/P project team \& NASA. HAO is a division of the National Center for Atmospheric Research, which is supported by the National Science Foundation; DOI: 10.5065/D64J0CXB). }
 \label{fig:1}
\end{center}
\end{figure}

Stellar observers are trying to find a counterpart to the observational signatures of CMEs that has been observed on the Sun. 
Here, we examine what observations have been made for the Sun and whether they can be applied to stellar observations.
It is important to note that while the Sun can be spatially resolved, stars cannot be spatially resolved. In the case of the Sun, most observations of CMEs have been made by direct imaging via Thomson scattering using a coronagraph \cite[(e.g., Howard 2011)]{2011JASTP..73.1242H}. With this method, statistical properties such as velocity and mass have been investigated and cataloged \cite[(e.g., Yashiro \& Gopalswamy 2009)]{2009IAUS..257..233Y}. However, it is very difficult to detect this radiation in stellar observations. Then, the following phenomena are considered to be promising for detecting the evidence of stellar CMEs (see Figure \ref{fig:1} for the summary of the solar observations).
Please also refer to an excellent review by \cite{2017IAUS..328..243O}.

{\underline{\it Prominence/filament eruptions seen as blue-shifted chromospheric lines}}.

Solar prominences and filaments are cool and dense plasma ($\sim$10$^4$ K and $\sim$10$^{10-11}$ cm$^{-3}$) floating in the corona \cite[(Gopalswamy et al. 2003; Shibata \& Magara 2011; Wood et al. 2016; Seki et al. 2021)]{2003ApJ...586..562G,2011LRSP....8....6S,2016ApJ...816...67W,2021EP&S...73...58S}.
Prominences and filaments are simply different in visibility and there is no essential physical difference. 
If the low-temperature plasma is floating outside the solar limb, it is observed as emission in the Balmer lines, and is called prominence. 
On the other hand, if the low-temperature plasma is floating inside the solar disk, they are observed as absorption lines and are called filaments (see Figure \ref{fig:1}).
Prominence/filament eruptions often happen in association with solar flares and sometimes become the core of CMEs \cite[(Gopalswamy et al. 2003; Wood et al. 2016)]{2003ApJ...586..562G,2016ApJ...816...67W}.
Although flares, CMEs, and filament eruptions are seen at different wavelengths, height, and timescale, they are only different aspects of a single phenomena triggered by the magnetic reconnection \cite[(Shibata \& Magara 2011)]{2011LRSP....8....6S}.
It should be noted that they are not necessarily all observed in association with each other, and therefore detecting only prominence/filament eruptions are not the conclusive evidence of CMEs (some prominence/filaments fail to be erupted after being lifted).
Prominence/filament eruptions are just lower-parts of the self-similarly expanding CMEs, and the velocities of prominence/filament eruptions are often less than the escape velocity and typically 2--8 times smaller than those of CMEs \cite[(Gopalswamy et al. 2003)]{2003ApJ...586..562G}.
\cite{2021EP&S...73...58S} provided a threshold to distinguish the solar filament eruptions with and without CMEs in the combinations of its length scale and velocity with high probability.
If we can estimate both values with stellar observations, it could be possible to speculate whether the stellar prominence/filament eruptions can evolve into stellar CMEs.

{\underline{\it Escaping coronal plasma seen as coronal dimming}}. 

Solar coronal dimming is a phenomenon where the coronal emission decreases after flares (see Figure \ref{fig:1}; \cite[Thompson et al. 2000]{2000GeoRL..27.1431T}).
These are thought to be a manifestation of coronal plasma, originally confined by closed magnetic fields, escaping from the solar surface due to the CMEs.
The signals of the coronal dimming have been reported in the Sun-as-a-star EUV spectral lines \cite[(Harra et al. 2016)]{2016SoPh..291.1761H}, and its application to stellar observations was suggested.
Later, \cite{2021NatAs...5..697V} extended this and find that solar coronal dimming is frequently associated with CMEs with a probability of more than 80 percent.
Therefore, coronal dimming is one of the most promising ways to find the occurrence of stellar CMEs, although it is not the direct emission from CMEs.

{\underline{\it Radio bursts}}.

Type-II radio bursts are radiated when the plasma and shockwave of CMEs are propagating through the interplanetary scape (see Figure 1).
If the CME is super-Alfvenic, shock waves are formed and radio emissions are generated by plasma oscillations. 
Since the plasma oscillation frequency (related to plasma density) varies with the height from the solar surface, it is known that the frequency drifts to the low-frequency direction.
The type-II radio bursts are known to have a strong association with CMEs \cite[(Gopalswamy 2006)]{2006GMS...165..207G}, but it requires a condition that the CME should be super-Alfvenic to generate the shock.
Type-IV radio bursts are also associated with flaring loops and/or CME flux ropes. 
Although the type-IV radio bursts are thought to have various emission sources and may not provide conclusive evidence of CMEs, they can be indicative of stellar CMEs.

\section{Indirect evidence of CMEs on M/K dwarfs, evolved stars, and binaries}

Based on the observations on the Sun in the Section 2, various methods have been proposed to detect stellar CMEs \cite[(see Korhonen et al. 2017; Osten \& Wolk 2017; Moschou et al. 2019)]{2017IAUS..328..198K,2017IAUS..328..243O,2019ApJ...877..105M}. 
Some signatures show nature similar to those of solar CMEs or prominence/filament eruptions, but other signatures have not been seen in the case of the Sun but are considered possible CME signatures.
Most of the observations are on M/K-type stars as described in this Section.
There has been only one report of a CME signature on a solar-type star (G-type star), and the recent discovery will be presented in the Section 4. 

{\underline{\it Blueshifted ``emission" of chromospheric lines}}.

Finding the Doppler shift of the chromospheric line has been one of the most successful tools to find stellar prominence/filament eruptions.
So far, blueshifted ``emission" components (sometimes called ``blue asymmetry") of chromospheric lines have been reported during Balmer-line/UV flares mostly on active M-/K-dwarfs \cite[(Houdebine et al. 1990; Gunn et al. 1994; Fuhrmeister et al. 2008; Leitzinger et al. 2011; Vida et al. 2016; Honda et al. 2018; Vida et al. 2019; Muheki et al. 2020; Maehara et al. 2021)]{1990A&A...238..249H,1994A&A...285..489G,2008A&A...487..293F,2011A&A...536A..62L,2016A&A...590A..11V,2018PASJ...70...62H,2019A&A...623A..49V,2020MNRAS.499.5047M,2021PASJ...73...44M}.
The blueshifted components may be evidence of stellar ``prominence" eruptions outside the stellar limb, that could be the core of stellar CMEs (see previous section). 
However, difficulty in interpreting these findings is that all blueshift components on M-/K-dwarfs are only seen as ``emission" asymmetric component of spectral lines.
In the case of solar flares, the flare-related emission at the footpoints sometimes shows blue asymmetry in the chromospheric line that is probably associated with evaporation \cite[(Tei et al.  2018)]{2018PASJ...70..100T}, and it has been difficult to distinguish between this evaporation-related cool upflows and prominence eruptions when the velocity is low.
Recently, \cite{2022MNRAS.513.6058L} suggested that this ``emission" may be because the stellar background emission components are quite weak on these cool stars and, unlike the Sun, the ``filament'' can be visible as emission even on the stellar disk (see also \cite[Odert et al. 2020]{2020MNRAS.494.3766O}).

It is true that this method has some ambiguity as mentioned above, but there is a strong advantage that it is possible to find its signature even from the ground and there have been hundreds of reports so far on cool stars (many of them are snapshot data, see \cite[Vida et al. 2019]{2019A&A...623A..49V}).
One the other hands, one of the significant disadvantages for these findings is that in many cases the velocity is only a few hundred km s$^{-1}$ or less, and it is not certain that it can evolve into CMEs eventually.
What this method can capture is prominence eruptions which are not necessarily direct evidence of CMEs.
In some cases, the velocities exceed the escape velocity \cite[(e.g., Houdebine et al. 1990; Vida et al. 2016)]{1990A&A...238..249H,2016A&A...590A..11V}, and these velocities have been a key to characterize the blueshift as probable evidence of stellar CMEs.

{\underline{\it X-ray, EUV, and FUV dimming}}. 

As explained above, coronal dimming is a very strong tool to constrain the occurrence of stellar CMEs.
\cite{2021NatAs...5..697V} reported post-flare coronal dimming in X-ray and EUV wavelength by using the archive dataset of stellar flare observations on cool stars, such as Proxima Cen (M-type star) and AB Dor (K-type star).
More recently, \cite{2022ApJ...936..170L} detected a possible post-flare FUV dimming of Fe XII 1349 {\AA} and Fe XXI 1354 {\AA} emission on $\epsilon$ Eri (K-type star) with the Hubble data.
These detections are considered to be promising indirect evidence of CMEs.
However, the number itself is still small and the detections depends on the definition of the ``quiescent'' flux level that can change in time.

One important point is that the depth and slope of the light curve of the dimming can be measured from the stellar light curves \cite[(Veronig et al. 2021)]{2021NatAs...5..697V} . According to solar studies, it is known that the mass of the CME can be estimated from the depth of the dimming, and the velocity can be estimated from its slope \cite[(Mason et al. 2016)]{2016ApJ...830...20M}. 
Although these methods have not yet been applied to stars, and modeling of stellar atmospheres is required for their application to stars, they are considered to be very promising for exploring the properties of stellar CMEs.

{\underline{\it Type-II and type-IV radio burst}}.  

The type-II radio bursts will be the most promising signature of the stellar CME shockwave, but no signature has been obtained so far, although previous studies have tried to find it \cite[(e.g., Crosley \& Osten 2018a\&b)]{2018ApJ...862..113C,2018ApJ...856...39C}. 
Recently, one detection of type-IV radio burst from the nearby M dwarf Proxima Centauri was reported and may be the evidence for a stellar CME \cite[(Zic et al. 2020)]{2020ApJ...905...23Z}.
The missing problem of the type-II radio burst may indicate less or no CMEs on the stars, but since other stars are thought to have different coronal conditions from the Sun, the frequency of radio bursts may be different than expected, or current radio telescopes may simply be lacking the sensitivity \cite[(Alvarado-G{\'o}mez et al. 2020)]{2020ApJ...895...47A}. 
It is also possible that CMEs can occur without shockwave \cite[(Mullan \& Paudel 2019)]{2019ApJ...873....1M}. 
Further observations with high sensitivity at low frequency is required in the future, through LOFAR, SKA, or a possible lunar observatory.

{\underline{\it X-ray blueshifted emission line}}. 

A signature of stellar CMEs on an evolved star HR 9024 was reported as a blueshifted emission of the cool X-ray O VIII line (4 MK) during a superflare \cite[(Argiroffi et al. 2019)]{2019NatAs...3..742A}.
The superflare has the blueshift with a velocity of 90 km s$^{-1}$ and the authors interpreted it as a slow CME (cf. the escape velocity is 220 km s$^{-1}$). 
The blueshifted components in X-ray range can be mainly emitted from the chromospheric evaporation in the case of solar flares, but the possibility was excluded by pointing out that the other hotter lines ($>$ 10 MK) do not show any blueshift at that time. 
This is also reported as a candidate for a stellar CME, although there is no way to verify it on the Sun, since no such observation has been made in X-rays so far on our Sun due to the lack of the instrument.

{\underline{\it X-ray absorption}}.

X-ray absorptions during stellar flares are also thought to be one possible signature of stellar CMEs.
The X-ray absorptions can represent the existence of plasma obscuring the flaring radiation, probably stellar CMEs or filament eruptions. 
\cite{1999A&A...350..900F} (and later \cite[Moschou et al. 2017]{2017ApJ...850..191M}) reported that a superflare on Algol shows the continuous absorption in X-rays gradually decaying with an inverse square law with time, and they suggested that this can be explained by expanding plasma above the star.
The other promising candidates are summarized by \cite{2019ApJ...877..105M}.

\section{Filament eruption and possible CME on solar-type star}

\begin{figure}[htb]
\begin{center}
 \includegraphics[width=5.in]{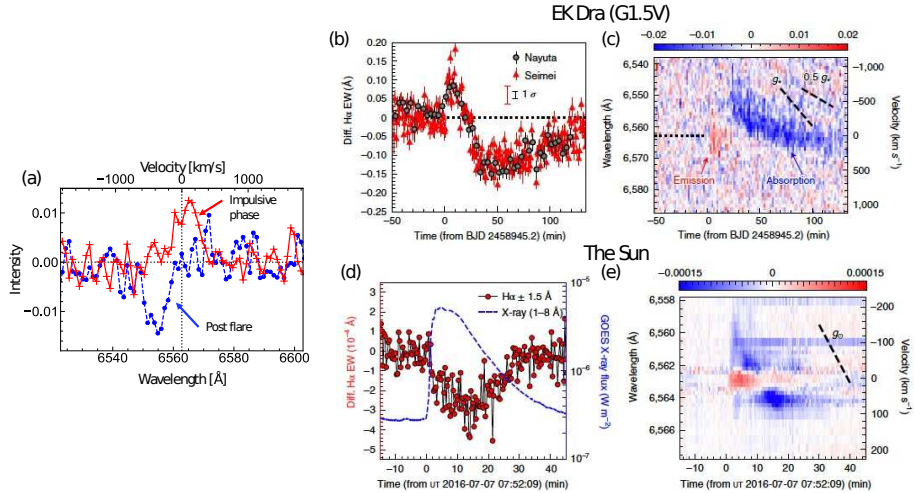} 
 \caption{Light curve and spectra of the superflare and filament eruption on the young solar-type star EK Dra (G1.5V) in comparison with the Sun-as-a-star analysis of a solar filament eruption (reproduced from \cite[Namekata et al. 2022a]{2022NatAs...6..241N}). (a) Typical pre-flare-subtracted spectra during the superflare and filament eruption on EK Dra. (b) Light curve of H$\alpha$ equivalent width of EK Dra (pre-flare level is subtracted). (c) Time evolution of pre-flare-subtracted spectra of H$\alpha$ line of EK Dra. (d) Light curve of H$\alpha$ equivalent width of the solar filament eruption. (e) Time evolution of pre-flare-subtracted spectra of H$\alpha$ line of the Sun. }
 \label{fig:2}
\end{center}
\end{figure}

The CME candidates presented in the previous section were mostly for cool stars such as M-type stars. Then, is there any evidence of CMEs on solar-type stars (G-dwarfs)?
In recent years, Kepler Space Telescope and TESS revealed the flare activity of solar-type stars and some active solar-type stars produce very large flare called superflares with the energy of $> 10^{33}$ erg \cite[(e.g., Maehara et al. 2012; Notsu et al. 2019; Okamoto et al. 2021; Namekata et al. 2022b)]{2012Natur.485..478M,2019ApJ...876...58N,2021ApJ...906...72O,2022ApJ...926L...5N}.
These have indicated a possibility that the ancient Sun could--and even the present-day Sun can--produce superflares, and possibly very large CMEs and geomagnetic storms. Therefore, the occurrence of CMEs on solar-type stars is important for the understandings of not only the habitability on the young Earth and Mars but also the possible extreme impact on the human civilization.

The question here is ``Can active solar-type stars produce stellar CMEs associated with superflares?'' and ``Are they really observable?".
According to Kepler and TESS studies, the frequency of detectable flares on old, slowly-rotating solar-type stars (age of several Gyr) is only once every a few thousand years, and it almost unobservable \cite[(e.g., Maehara et al. 2012; Notsu et al. 2019; Okamoto et al. 2021)]{2012Natur.485..478M,2019ApJ...876...58N,2021ApJ...906...72O}. 
For young solar-type stars (age of hundreds of Myr), however, the frequency of  detectable flares is about once every few days, which is realistically observable within a limited observing window \cite[(e.g., Namekata et al. 2022b)]{2022ApJ...926L...5N}.
Young solar-type stars are good target for the flare monitoring, even though we require long-term observations for weeks to detect one flare event.
The other difficulty in flare observation on solar-type star is the low amplitude of the flare radiation compared to the stellar luminosity. 
Let us discuss the case of optical observations of Balmer line profile.
The amplitude is estimated to be a few percent of stellar irradiance in optical wavelengths and the timescale is minutes to hours.
Considering that both high time resolution and high S/N is important to find this small transient, ground-based telescopes having a large aperture (e.g., 2-4 meter) is necessary.
In conclusion, we need long-term time-resolved observations of young solar-type stars with 2-4 m class telescopes to find one flare event on Sun-like star.

\begin{figure}[htb]
\begin{center}
 \includegraphics[width=5.in]{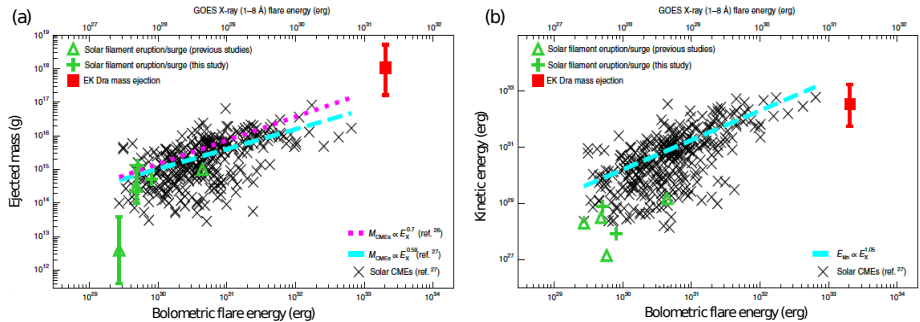} 
 \caption{Kinetic energy and ejected mass of solar and stellar CMEs/filament eruptions as a function of flare energy (reproduced from \cite[Namekata et al. 2022a]{2022NatAs...6..241N}).}
 \label{fig:3}
\end{center}
\end{figure}

Based on the estimation, we have performed a long-term time-resolved spectroscopic observations of H$\alpha$ line of a nearby young solar-type star EK Dra (G1.5V, age of $\sim$100 Myr).
We observed this star with the 3.8-m Seimei telescope in Japan \cite[(Kurita et al. 2020)]{2020PASJ...72...48K}, which has an advantage of a plenty of observational time ($\sim$30-50 nights for a challenging observational proposal).
As a result of our intensive survey, our optical spectroscopic observation reveals the first evidence for a stellar filament eruption associated with a superflare on a solar-type star \cite[(Namekata et al. 2022a)]{2022NatAs...6..241N}. 
The detected superflare on EK Dra has an energy of 2.0$\times$10$^{33}$ erg, and a blueshifted H$\alpha$ ``absorption" component with a high velocity of --510 km s$^{-1}$ was observed after the superflare (see Figure \ref{fig:2}).
The absorption signature is the conclusive evidence of the filament eruption on the star. 
We also performed the Sun-as-a-star analysis of solar filament eruptions and found that the temporal evolution in the spectra greatly resemble those of stellar filament eruptions (see Figure \ref{fig:2}d, e). 
This indicates that the picture of the filament eruptions on EK Dra is very similar to those on the Sun, although its energy scale and velocity are different.
Comparing this eruption with solar filament eruptions in terms of the length scale and velocity (see Section 1 and Seki et al. 2019), we suggest that a stellar CME did occur on EK Dra as a result of the superflare and filament eruption. 
The filament mass of 1.1$\times$10$^{18}$ g is ten times greater than those of the largest solar CMEs (see Figure \ref{fig:3}a).
Our first detection of filament eruption and possible CMEs on the young solar-type star provide a precious implications because it enables us to estimate how they affect the environment of young exoplanets and the young Earth.

One interesting point is that another superflare on EK Dra with much larger energy of 2.6$\times$10$^{34}$ erg did not show any signature of filament/prominence eruption \cite[(e.g., Namekata et al. 2022b)]{2022ApJ...926L...5N}.
How often does the eruptions occur on young solar-type stars? 
This is another important issue that needs to be clarified and will be discussed in Section 5.

\section{Conclusion and future direction}

In recent years, the number of reports on indirect evidence of stellar CMEs has become large. 
This is partly inspired by the advance of the exoplanet studies, but also largely due to technological advances and the accumulation of archival data.
The reported cases are not limited to a single wavelength, but are observed in X-ray, EUV, FUV, optical, and radio, although they are not simultaneously obtained.
Most of them have been reported on M-/K-dwarfs, but in recent years, detections on solar-type stars (G-dwarfs) have also been reported for the first time.
However, there has been no direct evidence of the stellar CMEs so far and more convincing detections are required. 
In addition, there are still few collaborations with numerical calculations, and there are also still few studies that investigate the impact of stellar CMEs on planetary atmospheres based on the observations of stellar CMEs. 
For further collaborations, we consider that the following points should be investigated in the future in this field.

{\underline{\it How can we see the stellar prominences?}}. 
Recent observations showed that the blueshift component can be seen in emission in cool M-dwarfs but in emission/absorption in the solar-type star and the Sun (see Section 3 and 4).
The dependence of the visibility of prominence/filament plasma on the stellar type and activity level should be investigated in the future. 
To investigate this, the radiative transfer calculations are necessary, and a prior study has been done by \cite{2022MNRAS.513.6058L}. 
Through these works, it is also important to propose a new method to investigate the physical quantities of stellar prominences.

{\underline{\it Are the stellar CMEs rare?}}.
Not all XUV/optical flares are accompanied by coronal dimming or Doppler shifts (see Section 4), and in fact the association rate of possible stellar CMEs against flares is low. Does this mean that the frequency of eruptive events is not very high? The frequency is a very important factor in determining how much they contribute to the impacts on the exoplanet habitability and stellar mass loss \cite[(Osten \& Wolk 2015)]{2015ApJ...809...79O}. We need to constrain this occurrence frequency of stellar CMEs with careful consideration of observational biases \cite[(Odert et al. 2020)]{2020MNRAS.494.3766O}.

{\underline{\it Why are kinetic energies small?}}.
The kinetic energy of the stellar CME candidates is smaller than extrapolated from the solar empirical laws of CMEs (see Figure \ref{fig:3}b). 
This may be due to the velocity difference between prominence eruptions and CMEs \cite[(Namekata et al. 2022a)]{2022NatAs...6..241N}, or because the velocity may be suppressed by the stellar magnetic field \cite[(Alvarado-G{\'o}mez et al. 2018)]{2018ApJ...862...93A}. To investigate the cause, it would be important to compare solar and stellar data under the same conditions (i.e., solar CME vs. stellar CME or solar prominence vs. stellar prominence). If the CME velocities can be determined from stellar XUV dimming, it may help to solve this issue.

{\underline{\it What is the more convincing evidence?}}.
Currently, no direct evidence of CME has been reported. 
One approach to examine more solid evidence is to simultaneously obtain indirect evidence of CMEs through multiple methods (e.g., coronal dimming and Doppler shift). 
This is expected to provide a more general view of stellar CMEs.



\end{document}